\documentclass[12pt,preprint]{emulateapj}

\usepackage[latin1]{inputenc}
\usepackage[T1]{fontenc}
\usepackage[english]{babel}
\usepackage{amssymb}
\usepackage[]{natbib}

\begin{document}
\title{The Excitation of N$_2$H$^+$ in Interstellar Molecular Clouds.
I - Models}
\author{F. Daniel\altaffilmark{1,2},J. Cernicharo\altaffilmark{1},M.-L. Dubernet\altaffilmark{2}}
\email{daniel@damir.iem.csic.es, cerni@damir.iem.csic.es,
marie-lise.dubernet@obspm.fr}
\altaffiltext{1}{Dept. Molecular and Infrared Astrophysics (DAMIR),
Consejo Superior de Investigaciones Cient\'{\i}ficas (CSIC),
C/ Serrano 121, 28006 Madrid. Spain}
\altaffiltext{2}{Observatoire de Paris-Meudon, LERMA UMR CNRS 8112, 5, Place
Jules Janssen, F-92195 Meudon Cedex, France.}

\begin{abstract}
We present LVG and non-local radiative transfer calculations involving the rotational 
and hyperfine structure of the spectrum of N$_2$H$^+$ with collisional rate coefficients 
recently derived by us. The goal of this study is to check the validity of the assumptions 
made to treat the hyperfine structure and to study the physical mechanisms leading to the 
observed hyperfine anomalies.

We find that the usual hypothesis of identical excitation temperatures for 
all hyperfine components of the $J$=1-0 transition is not correct within the range 
of densities existing in cold dense cores, i.e., a few 
10$^4$ $\textless$ n(H$_2$) $\textless$ a few 10$^6$ cm$^{-3}$. This is due to different 
radiative trapping effects in the hyperfine components. 
Moreover, within this range of densities and considering the typical abundance of 
N$_2$H$^+$, the total opacity of rotational lines has to be derived taking into account 
the hyperfine structure. The error made when only considering
the rotational energy structure can be as large as 100\%.

Using non-local models we find that, due to saturation, hyperfine anomalies appear 
as soon as the total opacity of the $J$=1-0 transition becomes larger than 
$\simeq$ 20. Radiative scattering in less dense
regions enhance these anomalies, and particularly, induce a differential increase of
the excitation temperatures of the hyperfine components. This process is more 
effective for the transitions with the highest opacities for which emerging 
intensities are also reduced by self-absorption effects.
These effects are not as critical as in HCO$^+$ or HCN, but
should be taken into account when interpreting the spatial extent of the
N$_2$H$^+$ emission in dark clouds.

\end{abstract}

\section{Introduction}
N$_2$H$^+$ was one of the first molecular
ions detected in interstellar space \citep{tha75}.
The $J$=1-0 line of this species has been extensively observed
toward cold dark clouds and protostellar cores to get some
estimates of the physical conditions of the gas (see, e.g., 
\citet{ber02,taf04,hot04,bel04,cas02}). These observations indicate that
N$_2$H$^+$ is a good tracer of the highest density regions of dark clouds. 
It seems that N$_2$H$^+$ is less depleted onto 
dust grain surfaces than CO and other molecular species. This is probably
related to the fact that N$_2$, the chemical mother species of N$_2$H$^+$,
is more volatile and condensates at lower temperatures than
carbon monoxide. In addition, the complex hyperfine structure of N$_2$H$^+$ 
increases the odds to have at least one optically thin hyperfine line component 
to probe the innermost regions of these clouds. Therefore, this species is in 
principle an interesting tool to study cold dark clouds. However, a drawback 
has been the lack of collisional rate coefficients between 
N$_2$H$^+$ and molecular hydrogen (or helium). The observational
data indicate some hyperfine intensity anomalies that could be due to
selective collisional processes or to radiative transfer effects (see
\citet{gon93} for the case the analogous case of HCN hyperfine intensities 
in dark clouds).

For current research, it is clear that astronomers need to know the state 
to state collisional rates of N$_2$H$^+$ with H$_2$ and He. This will 
be even more necessary for ALMA due to the much higher angular resolution and
higher sensitivity observations that it could provide of protostellar cores 
in several rotational transitions of N$_2$H$^+$ (up to $J$=9-8).

N$_2$H$^+$ has also been detected in warm molecular clouds (e.g. \citet{tur77}) 
where the lines are broader and very strong. In these objects only 
the hyperfine structure due to the external N-atom could be noticed as 
the splitting produced by the internal N-atom is lower than the intrinsic 
line width. Nevertheless, in order to correctly model the N$_2$H$^+$ intensities 
emerging from these clouds, astronomers need a complete set 
of state to state collisional rate coefficients for high temperatures.

A detailed study on molecular ions excitation was carried out by \citet{gre75}
and a set of collisional rate coefficients for N$_2$H$^+$ colliding with He was provided. 
In that work, the rate coefficients were given for transitions among N$_2$H$^+$ rotational 
energy levels. Recently, \citet{dan05} 
extended the previous study by computing a new set of collisional rate coefficients 
for transitions among hyperfine energy levels and using a new potential energy surface.
The range of kinetic temperatures is between 5-50 K 
and in a future paper, rate coefficients for temperatures up to 300 K will 
be provided.

The paper is organized as follows: Section \ref{spectroscopy} is devoted to the 
spectroscopy of N$_2$H$^+$. In section \ref{LVG} we present the results obtained with 
a Large Velocity Gradient (LVG) model to discuss the treatment of molecular hyperfine 
transitions. Comparisons with the cases of HCN and HCl are made. In section \ref{non-local} we 
present results from non-local radiative transfer models applied to N$_2$H$^+$ 
for different cloud structures. 

\section{Spectroscopy of N$_2$H$^+$}\label{spectroscopy}
\begin{figure}
\epsscale{0.6}
\plotone{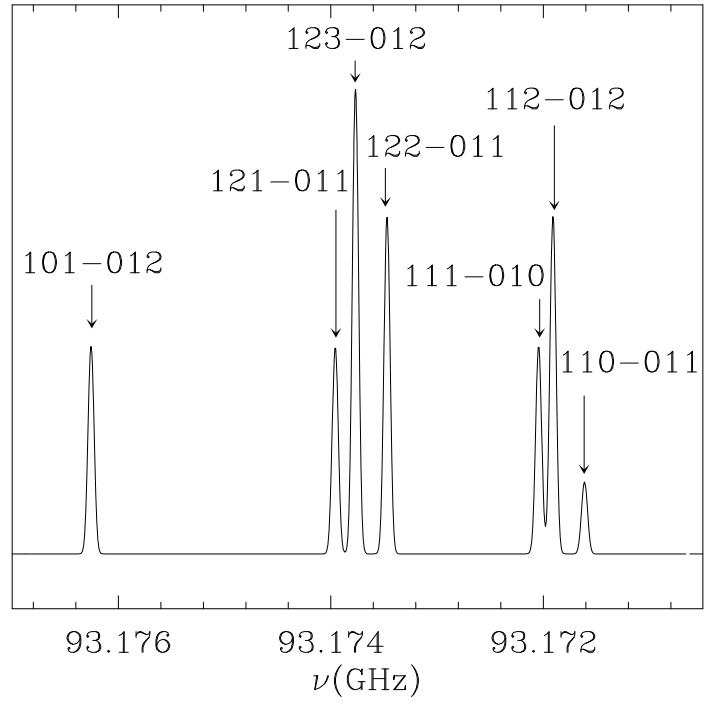}
\caption{
Hyperfine transitions associated to $J$=1-0 in the optically thin case.
Each resolved frequency 
is associated to multiple transitions among hyperfine levels (see text), and
the standard labeling of the lines is indicated.\label{f1}}
\end{figure}

\begin{figure}
\includegraphics[angle=270,scale=0.8]{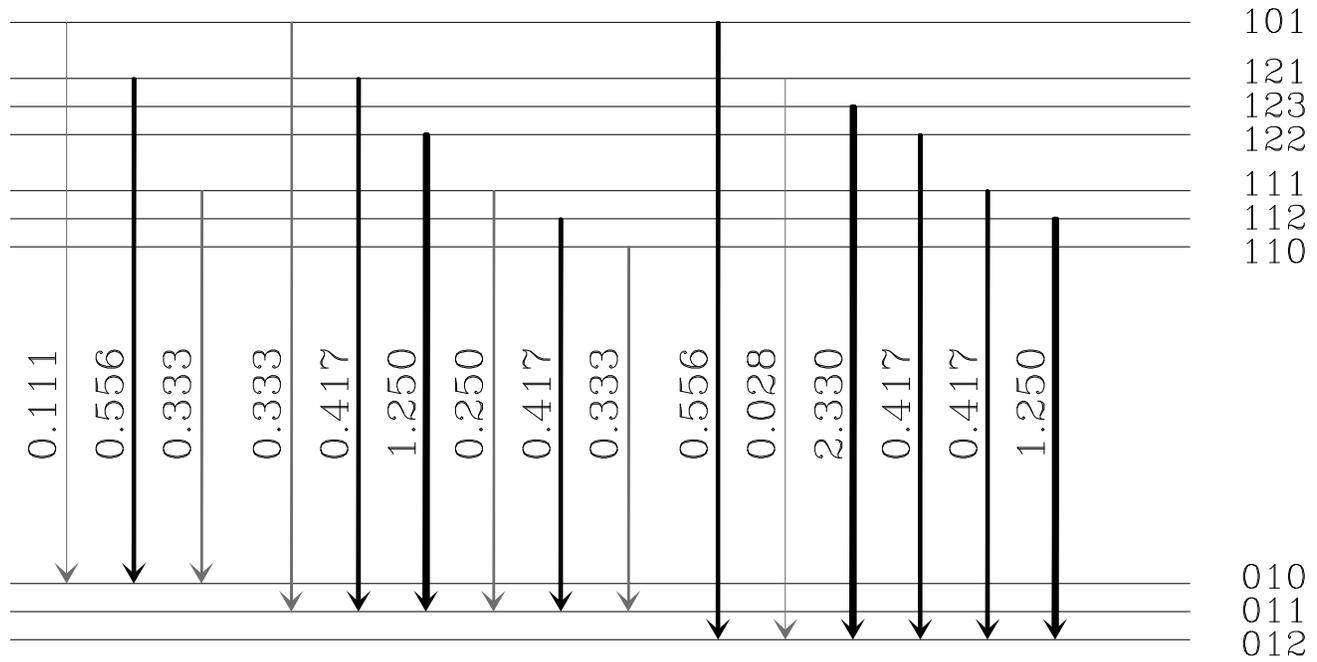}
\caption{
Line strengths of the 15 hyperfine components in $J$=1-0 transition. The thickness
of the lines indicate their relative weight compared to the others.
Lines strengths are normalized in such a way that, summing over all initial $J$=1 levels,
gives the degeneracy of the final $J$=0 levels, i.e. 5 for $JF_1F$=012,
3 for $JF_1F$=011 and 1 for $JF_1F$=010. Thus, the sum over all 15 transitions gives
 the total spin degeneracy.\label{f2}}
\end{figure}

The energy levels of N$_{2}$H$^{+}$ are characterized by the quantum
numbers $J$ (rotational quantum number), $F_{1}$ that results from the 
coupling of $\hat{J}$ with $\hat{I}_{1}$ ($\hat{F}_{1}=\hat{J}+\hat{I}_{1}$, 
where $I_{1}$=1 corresponds to the nuclear spin of the outer nitrogen), and $F$ 
($\hat{F}=\hat{F}_{1}+ \hat{I}_{2}$, where $I_{2}$=1 for the inner nitrogen). We have 
maintained the symbols used in \citet{dan04,dan05}, except for $J$. 
The external nitrogen nucleus induces the largest splitting since its coupling 
constants are larger than those of the internal nucleus. Following \citet{gordy}, 
the hyperfine energy levels can be found by diagonalizing the 
molecular Hamiltonian $H_{mol} = B \hat{J}^2 -D\hat{J}^4 + H_{coupling}$, 
where $B$ and $D$ are respectively the rotational and centrifugal 
distortion constants of the molecule and $H_{coupling}$ the effective 
nuclear coupling Hamiltonian. The Einstein coefficients ($A_{u \to l}$) are given 
by the equation: 

\begin{eqnarray}
\displaystyle
A_{JF_{1}F\to J'F'_{1}F'}= \frac{64\pi^{4}}{2hc^{3}}
\mu^{2}\nu_{JF_{1}F\to J'F'_{1}F'}^{3}
\times \frac{J'}{[F]}s_{JF_{1}F\to J'F'_{1}F'}
\end{eqnarray}

with the line strengths given by:

\begin{eqnarray}
s_{JF_{1}F\to J'F'_{1}F'}= [F_{1}F'_{1}FF']
\left\{
\begin{array}{ccc}
J & F_{1} & I_{1} \\
F'_{1} & J' & 1 \\
\end{array}
\right\}^{2}
\left\{
\begin{array}{ccc}
F_{1} & F & I_{2} \\
F' & F'_{1} & 1 \\
\end{array}
\right\}^{2}
\label{coeff_hyp}
\end{eqnarray}
where [x] stands for (2x+1) and \{.\} is the Wigner-6j
coefficient. The values adopted for rotational and coupling constants 
were provided by L. Dore (see \citet{dan04})
and where determined following the method described in \citet{cas95}. 
We assume a dipole moment for N$_2$H$^+$ of 3.4$\pm$0.2 D, as derived
experimentally by \citet{hav90}. It is in excellent
agreement with the value of 3.37 D derived theoretically by \citet{bot84}.
The resulting frequencies and line strengths for
the $J$=1-0 transition are given in Table \ref{table:spectroscopy}.
Figure \ref{f2} shows energy diagrams of the $J$=1 and $J$=0 hyperfine 
levels and indicates the line strengths of the hyperfine transitions. Figure 
\ref{f1} shows the $J$=1-0 hyperfine components together with their labeling.

Due to hyperfine interactions, there are 9 distinct energy levels for
$J \textgreater$1, 7 for $J$=1 and 3 for $J$=0. It is worth noting that the three energy
levels in $J$=0 are indistinguishable from a spectroscopic point of view, 
as their energy splitting is less than 10$^{-6}$ Hz.
Thus, although there are actually 15 allowed hyperfine transitions connecting $J$=1
to $J$=0, there are only 7 resolved features, usually labeled as: 
110-011, 112-012 (112-011, and 012),
111-010 (111-010, 011, and 012),
122-011 (122-011 and 012),
123-012, 121-011 (121-010, 011, and 012),
and 101-012 (101-010, 011 and 012). Hereafter, we call
hyperfine component each one of 15 transitions, and 
set of transitions is used to refer to each group among the seven groups 
of blended components, using the labeling indicated above.

\section{LVG models}\label{LVG}

\subsection{Effect of collisional rate coefficients}\label{rate_LVG}
\begin{figure}
\includegraphics[angle=270,scale=0.6]{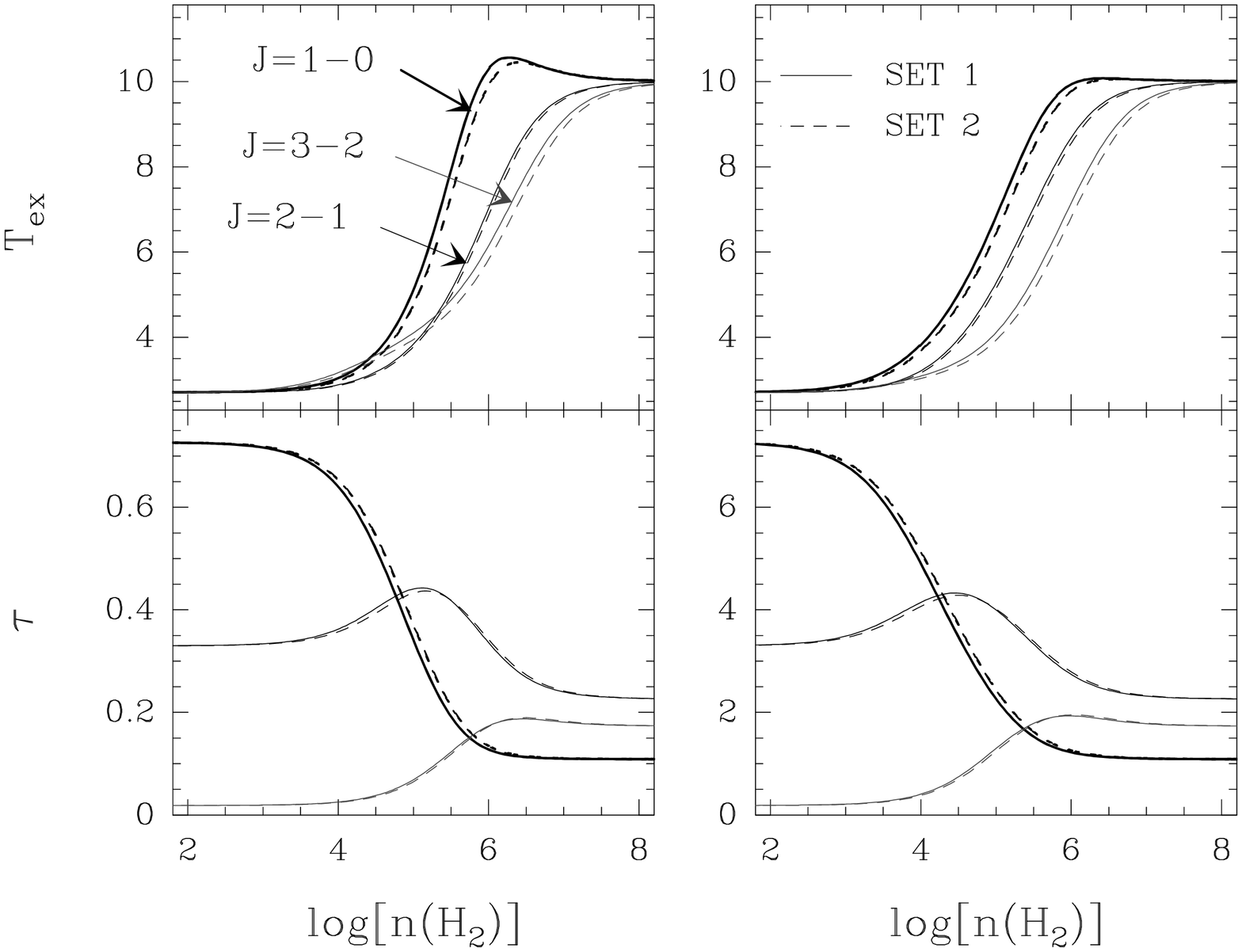}
\caption{Excitation temperatures (T$_{ex}$) and opacities ($\tau$) obtained 
  with two different sets of collisional rate coefficients and a LVG code,
  for a temperature of T = 10K. 
  SET 1 (solid lines) refers to the latest computed rate coefficients
  \citep{dan05} and SET 2 (dashed lines) to the previously available one \citep{gre75}.
  Left and right panels correspond to N$_2$H$^+$ column densities of
  respectively 10$^{12}$ and 10$^{13}$ cm$^{-2}$. A systematic
  velocity field of 1 km s$^{-1}$ has been assumed for the cloud.\label{f3}}
\end{figure}

To date, two sets of collisional rates are available for the rotational
structure of N$_2$H$^+$: \citet{gre75} and \citet{dan05}
\footnote{It was incorrectly stated, in \citet{dan04,dan05} that
the potential energy surface (PES) used by \citet{gre75} was calculated under the
electron-gas approximation. Indeed, as explained in \citet{gre75}, the PES was 
determined using the self-consistent field approach which gives accurate results for 
interacting systems where the charge-induced dipole term is prominent. This is actually
the case for the N$_2$H$^+$ - He system and explain the agreement between
the two sets of collisional rate coefficients for the transitions among the first 
rotational levels.}.
The differences between state to state rate coefficients
increase with both $\Delta J$ and $J$ as discussed in \citet{dan05}. In Figure 
\ref{f3} are compared opacities and excitation temperatures
obtained using the two different sets in the same LVG code. 
There is agreement within a few percent, and the main differences 
are mainly due to changes in critical densities. 
At 10 K, the rate coefficients for $\Delta J = 1$ agree within 20\% and the critical
densities for the radiative lines differ by a similar factor in the opposite way. 
In particular, the rate coefficient for the transition $J$=1-0 is 22\% higher using the 
results of \citet{dan05} and the derived critical density is 22\% smaller. In other words, 
the recently derived rate coefficients make this transition thermalized at slightly 
lower densities.

For densities below 10$^4$ cm$^{-3}$ the excitation temperature of the
$J$=1-0 line would be very close to the cosmic background temperature
due to the high dipole moment of N$_2$H$^+$ and the high critical densities
(see Figure \ref{f3}).
However,
for column densities as low as 10$^{12}$ cm$^{-2}$ (left panels of Figure
\ref{f3}),
the opacity of the $J$=1-0 and $J$=2-1 lines would be large enough to produce
significant absorption effects on the radiation emerging from the inner and
denser regions of the clouds. It is difficult to estimate from
emission measurements the abundance of N$_2$H$^+$ in these low density
regions. However, by observing the $J$=2-1 and $J$=3-2 lines it could be possible
to assess the effect of the external layers of the cloud on the
emerging profiles of the low-$J$ lines of N$_2$H$^+$, and thus to indirectly
determine the spatial repartition in N$_2$H$^+$ abundance.
This effect will be discussed in detail below.

\subsection{LVG treatment of hyperfine transitions}
\begin{figure}
\includegraphics[angle=270,scale=0.7]{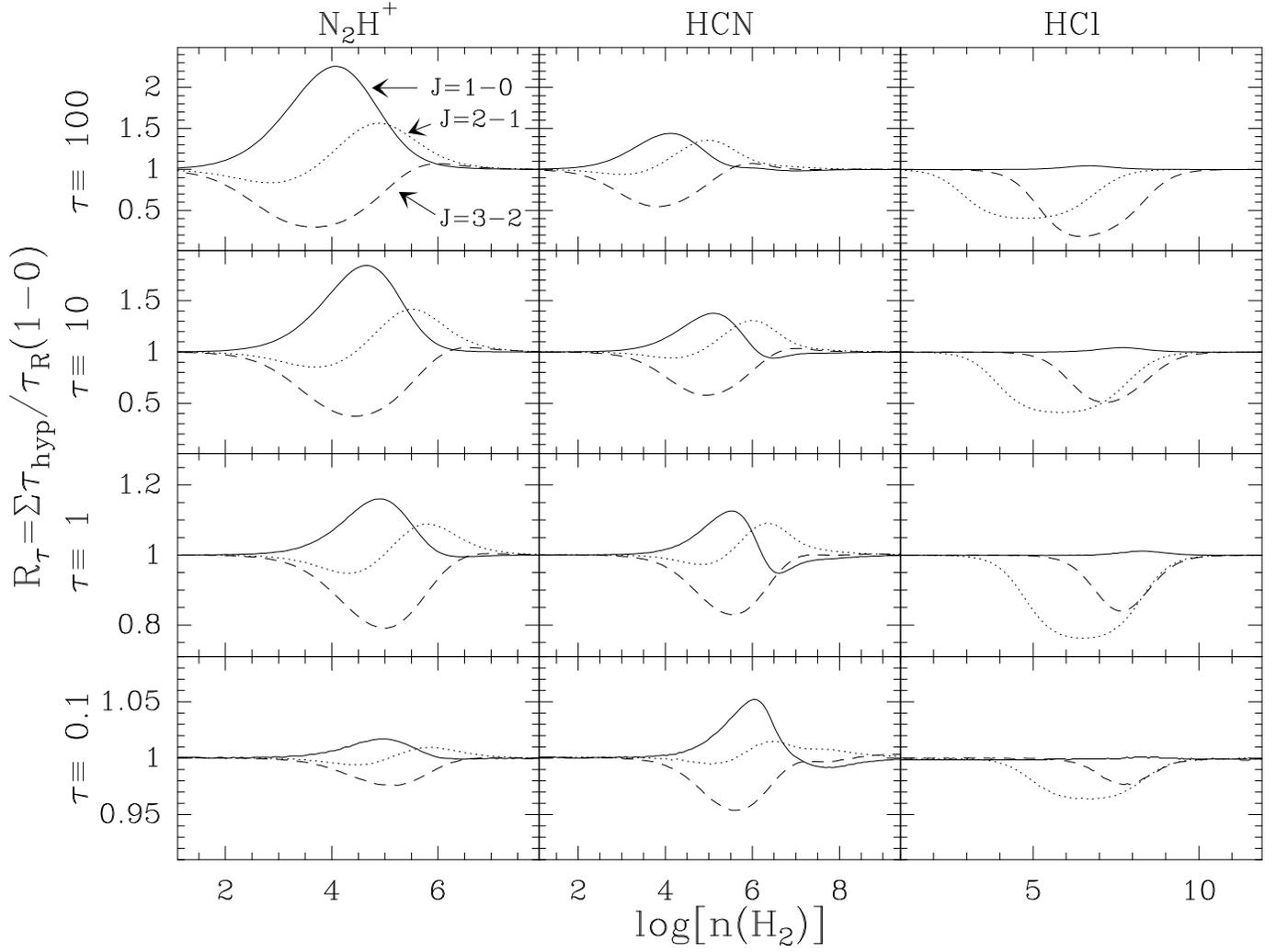}
\caption{Ratio between the opacities of the $J$=1-0 (solid), $J$=2-1 (dotted) and 
$J$=3-2 (dashed) rotational lines, $R_\tau$, determined with and without hyperfine
description for three molecules presenting hyperfine structure : N$_2$H$^+$ (left
panels), HCN (central panels), and HCl (right panels). 
For each molecule the column densities are such that, at T=10K, opacities 
of the $J$=1-0 lines are $\tau(1-0) \sim$ 0.1, 1, 10 and 100.
The column densities are respectively : N(N$_2$H$^+$) $\simeq$ 2.2 10$^{11}$, 2.3 10$^{12}$,
2.5 10$^{13}$ and 2.0 10$^{14}$ cm$^{-2}$/(km s$^{-1}$ pc$^{-1}$);
N(HCN) $\simeq$ 3.7 10$^{11}$, 3.0 10$^{12}$, 2.5 10$^{13}$, and 3.4 10$^{14}$
cm$^{-2}$/(km s$^{-1}$ pc$^{-1}$); N(HCl) $\simeq$ 8.5 10$^{11}$, 6.9 10$^{12}$, 7.3 10$^{13}$, and
7.7 10$^{14}$ cm$^{-2}$/(km s$^{-1}$ pc$^{-1}$).\label{f4}}
\end{figure}

The LVG model used for the simulations presented in this work 
takes into account all possible collisional and radiative
transitions among hyperfine levels. Nevertheless, it does not
account for local overlap, and particularly, the 15 transitions associated
with $J$=1-0 line are treated independently. 
In order to check the effect of the hyperfine structure,
we performed calculations with and without it for three species: 
HCl, HCN and N$_2$H$^+$. Figure \ref{f4} shows, for the three species,
 the ratio of the summed opacities of the hyperfine components, 
$\tau(1-0)$=$\Sigma\tau_i$, to the opacity of the unsplitted rotational 
transition, $\tau_R(1-0)$. This ratio will be referred to as 
$R_\tau$=$\tau(1-0)$/$\tau_{R}(1-0)$. The range of $\tau(1-0)$ explored is 0.1 
to 100. This figure reveals differences between the two approaches. 
Under LTE conditions (large volume density), or in radiative equilibrium with 
the cosmic radiation background
(low volume density), the two approaches lead to the same estimate of the 
total opacity because under these conditions the 
hyperfine levels are populated according to the statistical weights :
\begin{eqnarray}
n_{JF_1F}=\frac{[F]}{[JI_1I_2]}n_J
\end{eqnarray}
where [$J$$I_1$$I_2$] is the total number of hyperfine sub--magnetic levels
for a given rotational quantum number $J$.
Such a population scheme of hyperfine energy levels occurs when lines are 
optically thin or in the domains of low and high volume densities,
 as Einstein ($A_{u \to l}$) and collisional ($C_{u \to l}$) rate coefficients reduce similarly after 
summation over initial and final hyperfine levels: 
\begin{eqnarray}
& \displaystyle \sum_{F_1F_1'FF'}[F]C_{JF_1F\to J'F'_1F'}=[JI_1I_2]C_{J\to J'} \\
& \displaystyle \sum_{F_1F_1'FF'}[F]A_{JF_1F\to J'F'_1F'}=[JI_1I_2]A_{J\to J'} 
\end{eqnarray}

If n(H$_2$) $\sim$ $A_{J\to J'}/C_{J\to J'}$,
both collisional and radiative processes compete in populating the 
energy levels. Figure \ref{f4} shows that in this case the excitation 
processes are less effective to populate the first 
excited rotational levels if the hyperfine structure is considered. 
Thus, the total population of the fundamental level $J$=0 is
underestimated in a treatment that only includes the rotational
molecular structure. It gives rise to an underestimation of
the opacity of the $J$=1-0 and $J$=2-1 lines, and to
an overestimate of the opacity of the $J$=3-2 transition of
N$_2$H$^+$ and HCN. For HCl there is always an overestimate of 
the opacity.

When the hyperfine structure is taken into account,
both Einstein and collisional rate 
coefficients have the same order of magnitude than the corresponding values for the 
rotational structure.
Thus, the amount of radiative de--excitation, in the hyperfine
description, is globally larger for the rotational lines. Then,
this makes R$_{\tau}$ increase with the total opacity of the lines.
As opacity increases, the different Einstein coefficients of hyperfine 
lines lead to different excitation conditions for each hyperfine component, 
i.e., different excitation temperatures (see Figure \ref{f7}).
In addition, in an inhomogeneous cloud, hyperfine lines may be  
excited in regions where other lines still have low excitation temperatures.  
This effect is analyzed in section \ref{non-local} with the help of 
non-local radiative transfer models.

HCl behaves differently with respect to HCN and N$_2$H$^+$ due to its large rotational 
constant (B $\sim$ 313.0 GHz). The first excited
levels are hardly populated at low temperatures. We can roughly
estimate the number of rotational levels reached by collisions to be
$\sim$k$_B$T$/$hB. Thus, for HCl, most molecules are in the fundamental 
$J$=0 level for T$_K$=10 K and $R_\tau \simeq$ 1. Although some 
discrepancies do appear for 
$R_\tau$ in high-$J$ transitions, the opacities 
of these lines are so low at T$_K$=10 K that the emerging intensity
is negligible. At low temperatures, a hyperfine treatment of HCl is thus not
necessary.

HCN and N$_2$H$^+$ have similar rotational constants (44.3 GHz and 46.6 GHz
respectively). For these two species, the excited rotational
levels are efficiently pumped by both collisions and radiation. Hence, the 
hyperfine structure has to be considered to correctly derive column densities 
and abundances. For the $J$=1-0 transition of N$_2$H$^+$
the error induced in the opacity estimate by neglecting the hyperfine structure
varies from 20\% ($\tau \sim 1$) to 100\% ($\tau \sim 100$).

\subsection{$J$=1-0 Hyperfine Brightness Temperature Ratios}
Hereafter, we call T$^i_{ex}$ the excitation temperature of the hyperfine component i. 
LVG calculations show that there are two domains in the estimate of T$^i_{ex}$ for 
$J$=1-0. When opacities of individual components are low ($\tau_i$ $\ll$ 1), 
the excitation temperature is the same for all hyperfine
components. In this case, the opacities are proportional to line strengths 
s$_{i}$ : $\tau_i = s_i \tau(1-0$). This behavior is similar to what
is expected in the LTE regime. For higher opacities, photon 
trapping induces different behaviors for the hyperfine components.

In the LTE approximation, where T$_{ex}$ is assumed to be the same for all 
lines, the brightness temperature ratios vary monotonously with the rotational line opacity :

\begin{eqnarray}
\frac{T_{B}^{i*}}{T_{B}^{j*}} = \frac{1-e^{-\tau_i}}{1-e^{-\tau_j}} \label{eq:ratio}
\end{eqnarray}

\begin{figure}
\epsscale{0.7}
\plotone{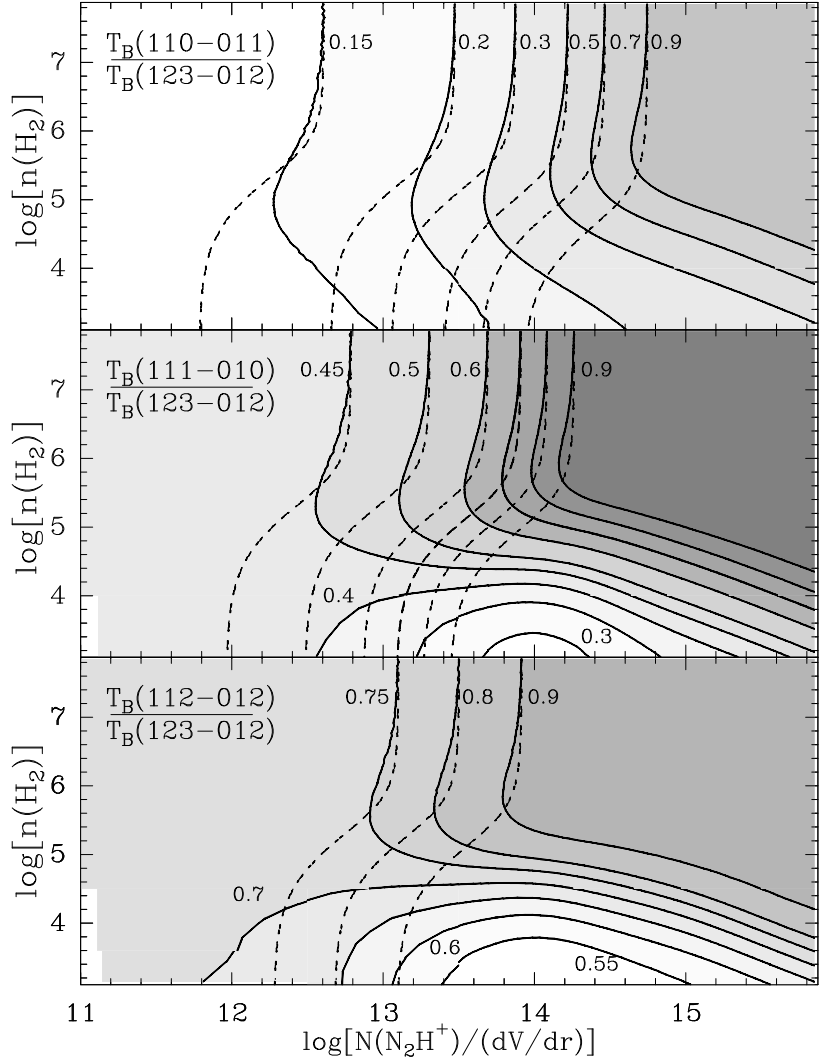}
\caption{Ratio of the brightness temperatures obtained from
  LTE calculations (long--dashed lines) to
  those from LVG calculation (continuous lines), 
  for a temperature of T=10K.
  The reference transition
  for both calculations is ($J$$F_1$$F$)=123-012.
  The abscissa corresponds
  to the N$_2$H$^+$ column density 
  (from 10$^{11}$ to 10$^{16}$ cm$^{-2}$/(km s$^{-1}$ pc$^{-1}$)) 
  and the ordinate is
  the volume density n(H$_2$) (10$^3$-10$^8$ cm$^{-3}$).
  Under LTE different transitions with the same initial
  quantum number $F$ have the same brightness
  temperature.
  From top to bottom the ratios of transitions
  with initial quantum number $F$ = 0,1, 2 are shown. These are
  compared to LVG results for the ratios
  T$_B$(110-011)/T$_B$(123-012),
  T$_B$(111-010)/T$_B$(123-012),
  and T$_B$(112-012)/T$_B$(123-012).
  Note that for high densities (n(H$_2$) $\textgreater$ 10 $^7$ cm $^{-3}$), i.e., thermalized
  conditions, dotted (LTE) and continuous (LVG) lines converge.
\label{f5}}
\end{figure}

Therefore, the brightness temperature ratios will change from the optically thin case, 
where they are given by the statistical weights 
 1 : 3 : 5 : 7, to unity in the optically thick case. Figure \ref{f5} shows
the hyperfine ratios, in LTE and non-LTE conditions, of the different hyperfine transitions 
with respect to 
the thickest one, i.e., $J$$F_1$$F$=123-012. As expected for high 
densities, the LVG calculations are coincident with the result of the LTE approximation. 
The LVG calculations show that in a large domain of the
(n(H$_2$), N(N$_2$H$^+$)) plane, i.e.  n(H$_2$) $\textless$ 10$^6$ cm$^{-3}$ 
and  N(N$_2$H$^+$) $\textgreater$ 10$^{12}$ cm$^{-2}$/(km s$^{-1}$ pc$^{-1}$), the
ratios are smaller than expected in the LTE approximation. Furthermore, 
ratios below the lower possible limit predicted under LTE conditions, 
i.e. 1/7, 3/7, 5/7, occur when hyperfine levels 
are not populated according to their statistical weights. 
Hence, fitting observational data using the hypothesis of identical excitation 
temperatures for all $J$=1-0 hyperfine components will lead to an 
erroneous determination of the total opacity. From our calculations it seems
that using this procedure underestimates $\tau$(1-0).

To discuss the behavior of the ratios, it is convenient
to define an averaged excitation temperature, $T_{ave}$, 
using LVG results for individual rotational  
lines, and their hyperfine components, as:
\begin{eqnarray}
T_{ave} = T_0 / ln \left( 1+\frac{T_0.\sum\tau_i}{\sum \tau_i.J_{\nu_i}(T_{ex}^{i})} \right)
\label{mean_Tex}
\end{eqnarray}
with T$_0$ = h$\nu$/k$_B$ and $\tau(1-0)=\sum \tau_i$.

\begin{figure}
\includegraphics[angle=270,scale=0.6]{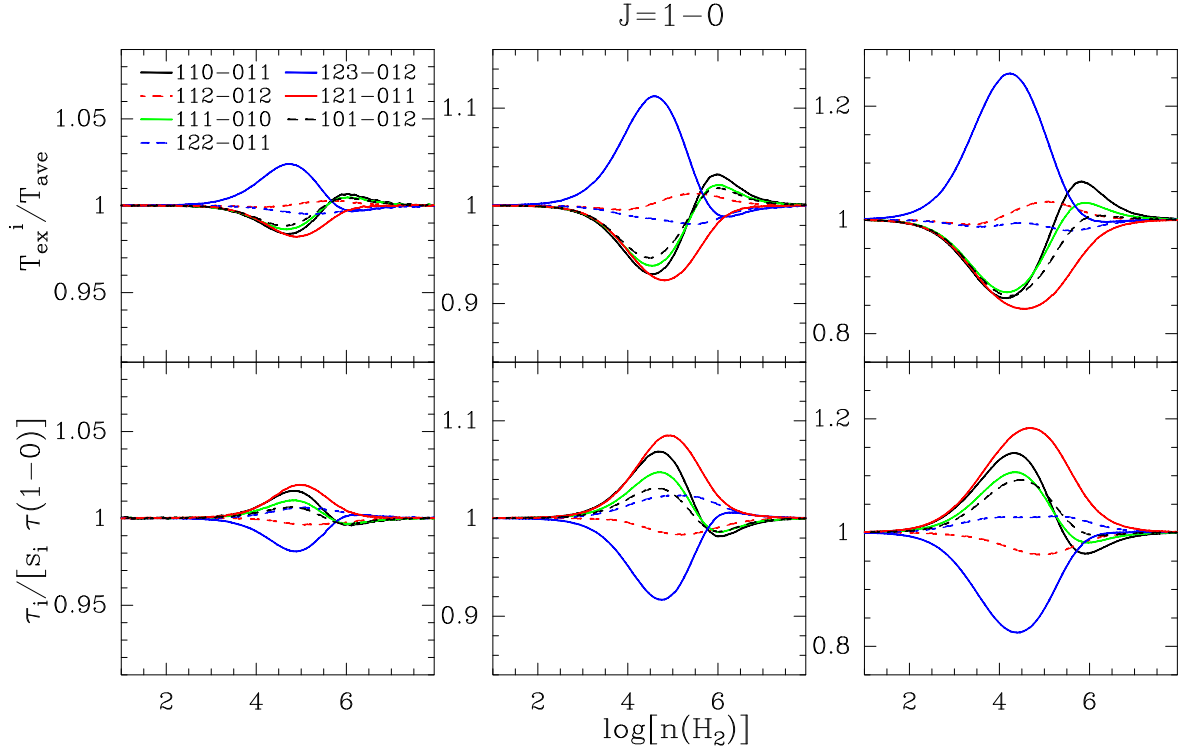}
\caption{Ratio of excitation temperatures and optical depths obtained
  in the LVG approximation to the corresponding quantities in the LTE
  limit. The temperature is T=10K and N$_2$H$^+$ column densities are 
  3.9 10$^{12}$ (left), 2.5 10$^{13}$ (center) 
  and 2.0 10$^{14}$ cm$^{-2}$/(km s$^{-1}$ pc$^{-1}$) (right column). 
  For each set of transitions, the opacities $\tau_{i}$ and the mean 
  excitation temperature T$_{ex}^{i}$ are obtained 
  by summing over lines with same frequencies and $\tau$(1-0) and T$_{ave}$ are 
  obtained by summing over all $J$=1-0 hyperfine lines (c.f. eq. \ref{mean_Tex}) . 
  Hyperfine set of transitions are referenced using the standard labeling (see Section 
  \ref{spectroscopy}). We see that the variation of the excitation temperatures is
  anti-correlated with the variation of the opacities.
\label{f6}}
\end{figure}

\begin{figure}
\includegraphics[angle=270,scale=0.6]{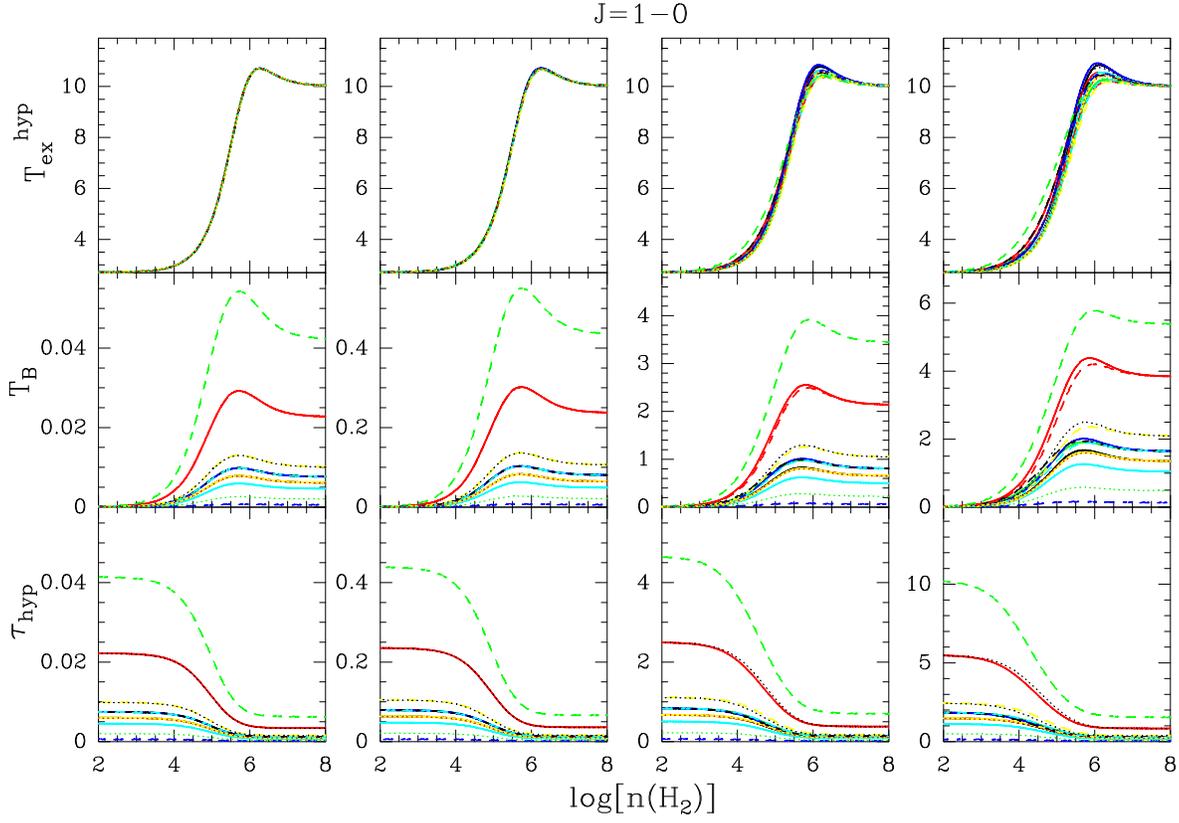}
\caption{ Excitation temperatures, opacities and brightness
  temperatures obtained in the LVG approximation, at T=10K, 
  for the 15 hyperfine
  components of the $J$=1-0 line. The columns, from left
  to right, respectively correspond to
   N(N$_2$H$^+$)=2.2 10$^{11}$, 2.3 10$^{12}$,
  2.5 10$^{13}$, and 2.0 10$^{14}$ cm$^{-2}$/(km s$^{-1}$ pc$^{-1}$). These
  column densities correspond to 
  $J$=1-0 total opacities of respectively $\tau(1-0) \sim$ 0.1, 1, 10, and 25.
  \label{f7}}
\end{figure}

Figure \ref{f6} shows the LVG excitation temperature
for all hyperfine components of the $J$=1-0 transition divided by $T_{ave}$.
This figure also shows the ratio of
individual hyperfine line opacities to the expected LTE opacities.
Figure \ref{f7} shows individual excitation temperatures 
and opacities for all hyperfine transitions. It is clear from these figures
that the effects are only important for large column densities, i.e.,
for large total opacities as discussed above. Non-LTE effects 
tend to reduce the spread of the opacities of the
different lines transferring opacity from the thickest to the
thinnest ones. 
For the hyperfine transitions associated to $J$=1-0, and
from Figures \ref{f6} and \ref{f7}, we find the
following trends :

\begin{itemize}
\item for $J$$F_1$$F$=123-012 : T$_{ex}^{i}$ $\textgreater$ T$_{ave}$ and
  $\tau_{i}$ $\textless$ $s_i \tau(1-0)$
\item for lines with initial quantum number $F$=2 : 
  T$_{ex}^{i}$ $\sim$ T$_{ave}$ and
  $\tau_{i}$ $\sim$ $s_i \tau(1-0)$
\item for lines with initial quantum number $F$=1 or $F$=0 :
  T$_{ex}^{i}$ $\textless$ T$_{ave}$ and
  $\tau_{i}$ $\textgreater$ $s_i \tau(1-0)$
\end{itemize}

We note that the variations of excitation temperatures and opacities
for the different hyperfine components are anti-correlated, i.e., 
an increase of $T_{ex}^i$ is accompanied by a decrease of $\tau_i$,
and inversely (see Figure \ref{f6}).
This fact suggests that non-LTE effects mainly induce variations
in the population of the $J$$F_1$$F$=012 level (the lower energy level in
the reference hyperfine transition) compared to the LTE values.
As for moderate and large opacities the brightness temperatures $T_B^{i*}$, 
are less sensitive than $T_{ex}^i$ to variations in $\tau_i$, the non-LTE 
effects tend to increase $T_B^*$ for the $J$$F_1$$F$=123-012 line and induce a decrease 
of $T_B^*$ for lines with initial quantum number $F$=0 or $F$=1. Thus, for a given
H$_2$ density and N$_2$H$^+$ abundance, ratios derived in the LVG 
approximation are smaller than those obtained under the LTE approximation
(see Figure \ref{f5}).

\section{Non-local Radiative Transfer for N$_2$H$^+$}\label{non-local}
The LVG calculations discussed in the previous section could be a 
reasonable approximation
to the emerging N$_2$H$^+$ intensities from dark clouds when line
opacities are low. However, as the line strengths of the hyperfine 
components are different, the density structure of the cloud, radiation scattering
and/or radiative coupling between different cloud regions could affect 
the population of N$_2$H$^+$ energy levels. 
In order to check the validity of the LVG approximation and the different assumptions 
made earlier in this paper to interpret the N$_2$H$^+$ $J$=1-0 transition, 
we carried out non-local calculations using the code developed and described by 
\citet{gon93}. For the opacity range considered in this work the code provides 
reliable and fast convergence.

We assume a cloud at 160 pc with an angular diameter of 30''
which corresponds to a radius of 3.6 10$^{16}$ cm (i.e. 0.023 pc), and we 
consider three different sets of models.
In the first one we consider a core of uniform density.
In the second set the central core is surrounded by an envelope with
size 3 and 6 times the core size. Finally, the last set corresponds to
a collapsing cloud. In all models the turbulence velocity was varied
from 0 to 0.4 km s$^{-1}$ by step of 0.1 km s$^{-1}$. Finally,
we adopt a kinetic temperature of 10 K in all models.

In all figures related to this section the ordinate
scale is antenna temperature 
obtained by convolving the cloud brightness temperature with the beam
of the 30-m IRAM radio telescope as follows: 
half power beam widths of 27'', 13.5''
and 9'', beam efficiencies of 0.76, 0.59 and 0.42, 
and error beams of 350'', 220'', and 160'' for the
$J$=1-0, 2-1 and 3-2 lines respectively. The error beam
is always larger than the size of the dense regions and thus
the energy entering the telescope radiation pattern through the
error beam just accounts for a few percent of the total intensity 
at most. Moreover,
the contribution to $J$=1-0 from the 
extended envelopes surrounding the cores is fully taken into account
by considering the convolution with the beam of the telescope
(main beam and error lobe). The only plots for which the
emerging intensity has not been convolved with the beam correspond
to the two lowest right panels of Figure \ref{f10}. In these cases,
we aim to show the influence of the high density region on the excitation
of the molecules in the low density envelope and this would have been 
hidden by performing the convolution with the telescope beam pattern.

\subsection{Uniform density cores}
\begin{figure}
\includegraphics[angle=270,scale=0.6]{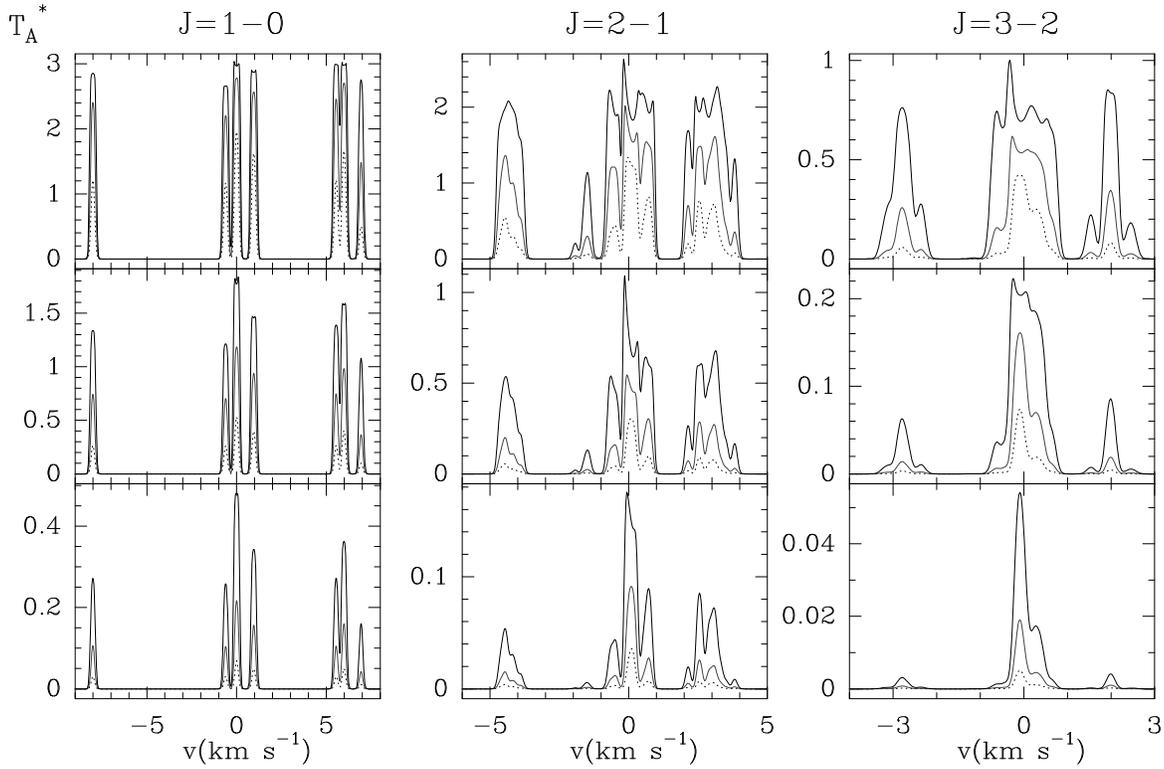}
\caption{
Emerging intensities for the $J$=1-0 (left panels), $J$=2-1 (central
panels) and $J$=3-2 (right panels) lines of N$_2$H$^+$. The
cloud has an uniform volume density, a radius of
3.6 10$^{16}$ cm, a turbulent velocity of 0.1 km s$^{-1}$, and a
kinetic temperature of 10 K.
In each box the line profiles are shown for N$_2$H$^+$
abundances of 4 10$^{-10}$ (dotted), 1.6 10$^{-9}$ (grey),
 and 6.4 10$^{-9}$ (solid). From top to bottom the three panels 
for each transition correspond to volume densities, n(H$_2$), of
2.5 10$^4$ (bottom), 10$^5$, and 4 10$^5$ cm$^{-3}$ (top). 
\label{f8}}
\end{figure}

For this set of models the N$_2$H$^+$ abundance has been varied
from 4 10$^{-10}$ to 6.4 10$^{-9}$, and the volume density 
from 2.5 10$^4$ cm$^{-3}$ to 4 10$^5$ cm$^{-3}$, by multiplicative steps of 4.
Figure \ref{f8} shows the emerging line profiles for the
$J$=1-0, 2-1 and 3-2 transitions for all computed densities n(H$_2$) 
and abundances X(N$_2$H$^+$). As expected, the
brightness temperature increases with n(H$_2$) due to 
collisional excitation, leading to the thermalization of the
the $J$=1-0 hyperfine components. 
We see that the brightness temperatures of the $J$=1-0, 2-1, 
and 3-2 transitions are more sensitive to a variation of n(H$_2$) in
the range 2.5 10$^4$cm$^{-3}\textless$ n(H$_2$) $\textless$4 10$^5$cm$^{-3}$ than to an
increase of X(N$_2$H$^+$). This can be understood by looking at figure
\ref{f7}: in this range of n(H$_2$) the excitation temperatures increase rapidly. 
Moreover, for $J$=1-0 total opacities below 20, the excitation
temperatures depend essentially on the volume density
as radiative trapping effects are negligible and do not induce 
significant departure for the different $T_{ex}^i$ from a single value.
Consequently, in the optically thin case and for a given volume density,
the line intensities are proportional to X(N$_2$H$^+$) 
(raw corresponding to n(H$_2$) = 2.5 10$^4$ cm$^{-3}$ on  Figure \ref{f8}). 
Radiative trapping and line saturation only become important for large 
N$_2$H$^+$ column densities. On Figure \ref{f8}, for the J=1-0 line, 
and on the panel corresponding to n(H$_2$)=10$^5$ cm$^{-3}$, we see that increasing
the abundance from 1.6 10$^{-9}$ to 6.4 10$^{-9}$ modify the ratio between the 
the $J$$F_1$$F$ = 101-012 and 121-011 sets of transitions. This change is due to radiative
trapping and entails variations in the $T_{ex}^i$ for the involved hyperfine components.
Thus for large opacities, i.e. $\tau$(1-0) $\sim$ 20, the excitation temperatures 
start to be sensitive to X(N$_2$H$^+$).

One of the main results that can be derived from Figure \ref{f8} is that
hyperfine intensity anomalies appear when increasing the N$_2$H$^+$ abundance. This 
is similar to the effect found in HCN and discussed
by \citet{gon93}. For total opacities in the $J$=1-0 transition above 10,
differential radiative trapping across the cloud starts to affect
the relative intensities of the hyperfine components.
Moreover, in an inhomogeneous cloud we could expect to be sensitive to different spatial extents 
for each hyperfine component (see \citet{cer87, gon93}).
This is even more striking when considering transitions with the same
initial quantum number $F$, expected to have the same brightness
temperature in the LTE limit.
The most affected hyperfine set of components is the 121-011 set
which indeed appears weaker than the 101-012 and 111-010 sets 
(see section 2). A low brightness temperature for this set has
already been observed (see e.g. \citet{lee99,cas95}).
We find the general trend $T_B$(111-010) $\gtrsim$ $T_B$(101-012) $\textgreater$
$T_B$(121-011) and $T_B$(112-012) $\gtrsim$ $T_B$(122-011). 
This behavior originates from differential non-LTE effects in
the hyperfine components of the $J$=0 $F_1$=1 levels ($F$=0,1,2).
This is not surprising as the three transitions
from $J$=1 $F$=1 ($F_1$=0,1,2) have different line strengths
(see Table \ref{table:spectroscopy}).
The 121-011 transition mostly depends on the $J$$F_1$$F$=010 level and is independent of
$J$$F_1$$F$=012 while the 111-010 and 101-012 sets
mainly depend on $J$$F_1$$F$=012. Thus, the 121-011 set weakness
compared to the other two must be related to an
underpopulation of the $J$$F_1$$F$=012 level that 
produces higher excitation temperatures for the hyperfine lines
involving it. As shown in Figure \ref{f8} the effect
appears for high abundances, i.e., large opacities, while for
low abundances the three sets have the same intensities. Hence, it seems
that it is not related to a collisional effect but to a radiative one.
The sum of the line strengths for lines connecting the $J$=1 to the
$J$=0 level is larger for $JF_1F$=012 than for the others. Thus, for large opacities
this level will be more affected by radiative trapping than the others.

\begin{figure}
\epsscale{0.6}
\plotone{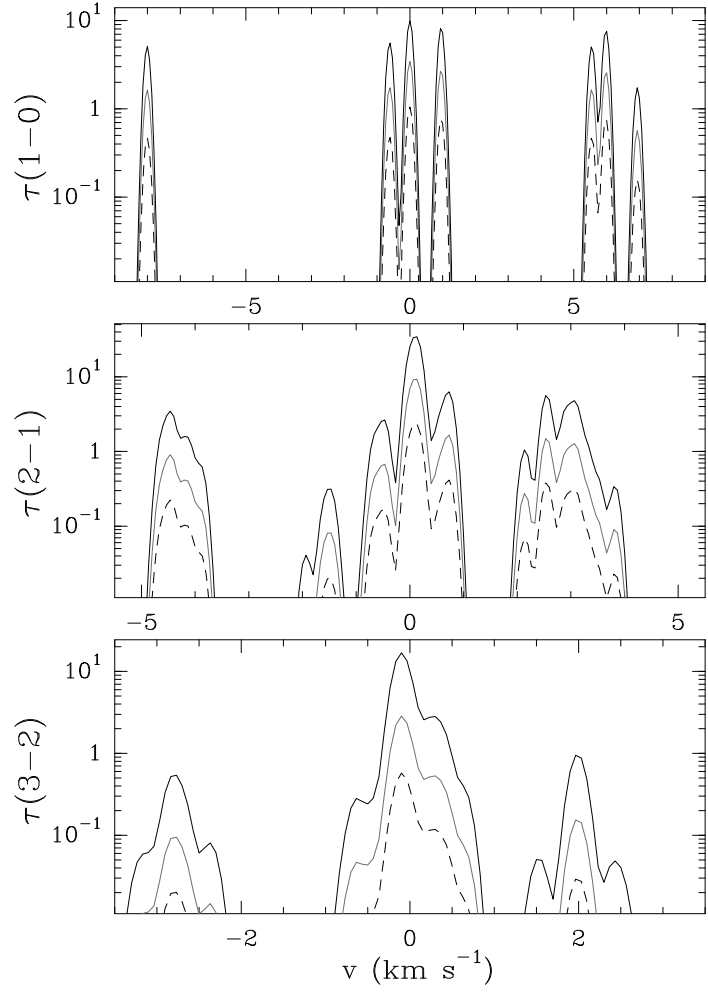}
\caption{Opacities for the $J$=1-0 (top), $J$=2-1 
and $J$=3-2 (bottom) lines of N$_2$H$^+$. The
cloud has a uniform volume density of 10$^5$cm$^{-3}$, a radius of
3.6 10$^{16}$ cm, a turbulent velocity of 0.1 km s$^{-1}$, and a
kinetic temperature of 10 K. N$_2$H$^+$ abundance is
4 10$^{-10}$ (dashed), 1.6 10$^{-9}$ (gray) and 6.4 10$^{-9}$ (black).
\label{f9}}
\end{figure}

For the $J$=2-1 and $J$=3-2 lines the effect of the abundance on the hyperfine
intensities is similar. Large abundances lead to larger
opacities, differential radiative trapping between the hyperfine components,
and to anomalies in the observed line intensities. For these levels,
line overlap is really important as the hyperfine components are closer 
in frequency and should be taken into account in a detailed modeling. 
In collapsing clouds the velocity field could
allow to have radiative coupling across the cloud
between hyperfine components with different frequencies.
This effect has been found to be extremely
important in molecules such as HCN \citep{gon93}.
In the models shown in Figure \ref{f8} we only consider the effects
of opacity for each line without taking into account population transfer
between levels due to radiative connection between them. Hence, the
effects shown in this figure for large abundances are purely due to the
line opacities. For the $J$=2-1 and 3-2 lines, the self absorption induced by the
external layers of the cloud is extremely important. When mapping clouds, 
the $J$=2-1 and 3-2 transitions have little intensity
outside the central position. However, although the excitation
temperature of their hyperfine components could be low for offset
positions, their line opacities
are still large enough to produce significant effects on the emerging
profiles toward the densest regions.
For example, the model in Figure \ref{f8}
with n(H$_2$) = 2.5 10$^4$ cm$^{-3}$ has low emission for all
lines. However, the opacities are large enough for the strongest hyperfine
components of the $J$=1-0 and $J$=2-1 lines to produce the
above mentioned effects. Figure \ref{f9} shows the
opacities of the lines for n(H$_2$)=10$^5$ cm$^{-3}$ which roughly
agree with those found in the case n(H$_2$)=2.5 10$^4$ cm$^{-3}$  multiplied
by a factor 4. We see that the opacity of $J$=2-1 will be even larger
than that of $J$=1-0, and of the order of 0.75/7.5 for the strongest
component and for the lowest/largest abundance of N$_2$H$^+$.
As shown in Figure \ref{f7}
the excitation temperatures of the  $J$=1-0 hyperfine components
vary strongly in the range 10$^4$ $\textless$
n(H$_2$) $\textless$ 10$^6$ cm$^{-3}$. The same will occur for the $J$=2-1 and 3-2
hyperfine components. Consequently,
strong
absorption will be produced in the low density layers of a cloud with
a marked density structure.

\subsection{Core/Envelope Clouds}
\begin{figure}
\includegraphics[angle=270,scale=0.6]{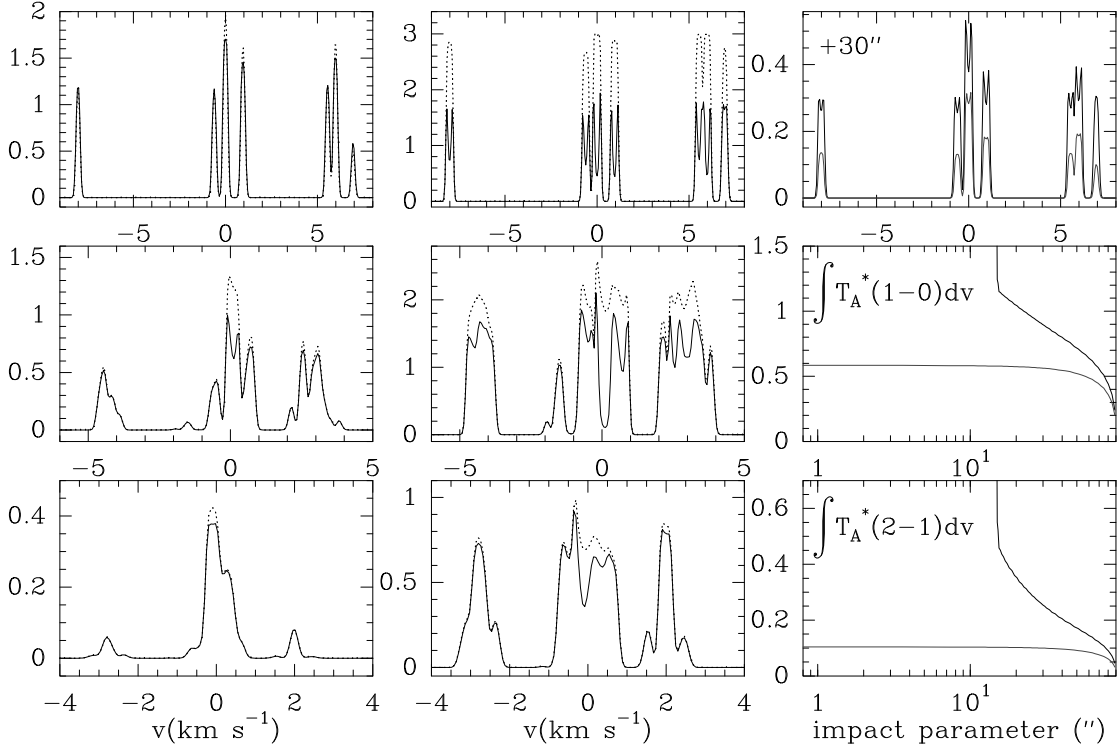}
\caption{
Emerging profiles for the $J$=1-0, $J$=2-1 and $J$=3-2 lines of N$_2$H$^+$
for a core/envelope geometry (left and central panels).
The core has a diameter
of 30'' and a density of 4 10$^5$ cm$^{-3}$. The envelope has
a diameter of 90''/180'' and a density of 5 10$^4$/5 10$^3$ cm$^{-3}$
for the left and central panels respectively.
The N$_2$H$^+$ abundance is 4/64 10$^{-10}$
(left/central panels). The kinetic temperature is 10 K and the turbulent velocity is
0.1 km s$^{-1}$ in the core and the envelope. Dotted lines correspond
to transitions emerging from the core alone while solid lines
correspond to those arising from the core+envelope system. The right
panels show the emerging spectrum at an offset position of 30'' (pure
envelope emission corresponds to the gray curve). The effect of the 
radiative excitation in the envelope due to the photons arising from 
the core is clearly seen in the two lowest right panels.
\label{f10}}
\end{figure}

We have finally explored more complex structures for the cloud to check 
the effects quoted above. 
Figure \ref{f10} shows the results for two different
cases. Left and central panels correspond to a simple core+envelope
structure. For the line profiles shown in
the left panels (T$_K$=10 K, n(H$_2$)$_{core}$=
4 10$^5$ cm$^{-3}$, n(H$_2$)$_{env}$=4 10$^4$ cm$^{-3}$,
X(N$_2$H$^+$)$_{core/env}$=4 10$^{-10}$, R$_{core}$=3.6 10$^{16}$ cm,
R$_{env}$=2 R$_{core}$), 
the opacity in the envelope
for the $J$=1-0 line is very small. The emerging profile is practically coincident with 
the one arising from the core. However, the $J$=2-1 line is optically thick
in the envelope and important absorption is caused to the emission 
from the core. As the opacity of the different hyperfine components
depends on the line strength, weak lines are not affected (components
at extreme negative and positive velocities), while the central
components are much more affected. The $J$=3-2 lines are practically not
affected by the envelope.

A different situation occurs if the abundance of N$_2$H$^+$
and/or the physical size of the envelope increase. 
The line profiles reported in
the central panels of Figure \ref{f10}
(T$_K$=10 K, n(H$_2$)$_{core}$=
4 10$^5$ cm$^{-3}$, n(H$_2$)$_{env}$=5 10$^3$ cm$^{-3}$,
X(N$_2$H$^+$)$_{core/env}$=6.4 10$^{-9}$, R$_{core}$=3.6 10$^{16}$ cm,
R$_{env}$=6 R$_{core}$) show that the strongest hyperfine components
of the $J$=1-0 line are strongly affected by absorption
(the effect is similar to the one obtained in the models shown in the previous
section, see Figure \ref{f8}, where absorption occurs
in the most external core layers where the excitation temperature 
has decreased with respect to the center).
For the $J$=2-1 hyperfine lines the presence of the envelope affect
dramatically the emerging intensities and the shape of the line. 
The $J$=3-2 line is less affected since the density in the envelope
is not large enough to pump the $J$=2 and $J$=3 levels efficiently. The core
also has an effect on the excitation temperatures of the
N$_2$H$^+$ lines in the envelope.  The right panels of Figure
\ref{f10} show the emerging profile from the envelope
at a position located 30'' away from the center. The top right panel shows the
comparison of the emission from the envelope alone and the emission obtained
when the core is also present and excites radiatively the N$_2$H$^+$ molecules
of the envelope. The intensity of the $J$=1-0 line has increased by a factor $\sim$ 2.
The other two right panels show the radial intensity distribution
(not convolved with the telescope beam) for the $J$=1-0 and
$J$=2-1 lines (integrated intensity over all hyperfine components). The
"heating effect" is clearly seen as a significant
increase of the integrated line intensity even at distances of 3-4 times the radius
of the core. These effects should be taken into account when interpreting
the radial distribution of N$_2$H$^+$ from observations of the $J$=1-0 line (which is much
easier to detect than the high-$J$ lines in these low density regions).

\subsection{Collapsing Clouds}
\begin{figure}
\includegraphics[angle=270,scale=0.6]{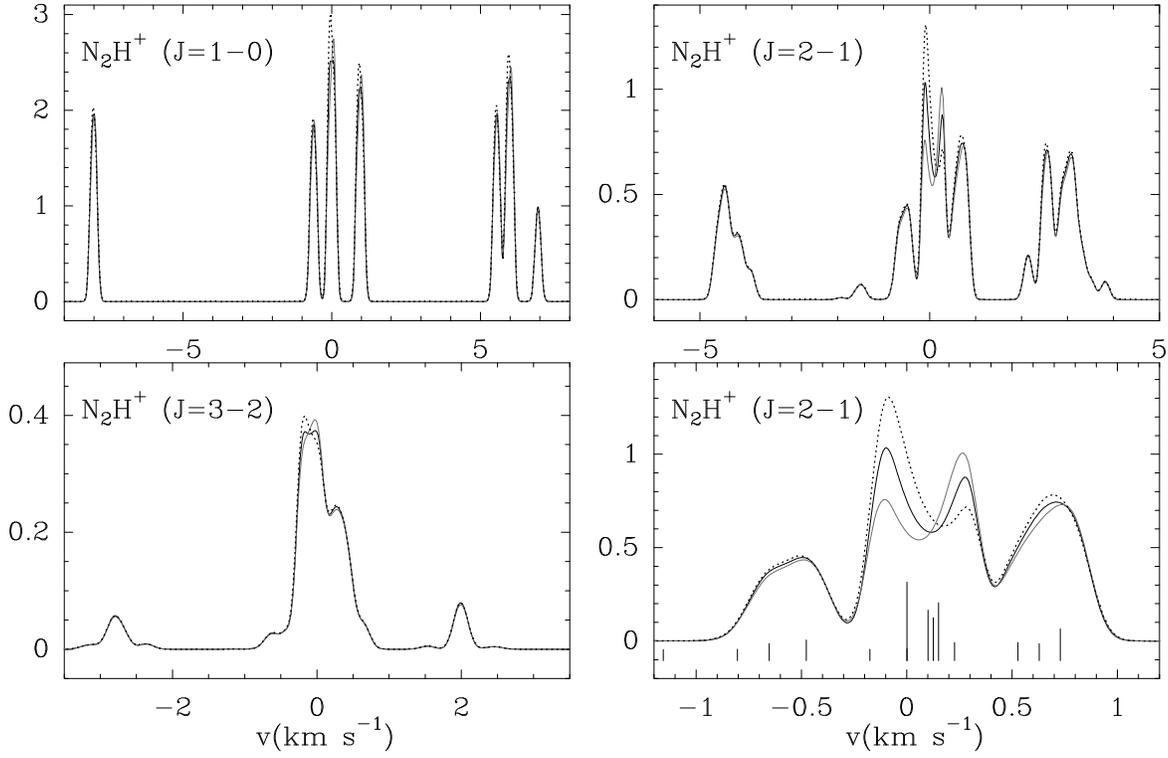}
\caption{Emerging profiles for a cloud with different velocity
fields: static (black line), increasing linear function (gray)
and  decreasing linear function (dotted). The lower right panel shows
the central lines of the $J$=2-1 line, with the positions of the 
individual hyperfine components indicated by lines proportional to their
line strength.
\label{f11}}
\end{figure}

Over the past decades, mm continuum observations and star count analysis \citep{war94}
have revealed that the density structure of cold dark clouds consists of an inner region 
of nearly uniform density and of an outer region where the density decreases as 
r$^{-p}$ with p $\sim$ 2.0-2.5. 
In this section, we use this type of density profile with a power 
law index p = 2 in order to test the influence of the 
velocity field on emerging spectra. 
Figure \ref{f11} shows the profiles
of the $J$=1-0, 2-1 and 3-2 lines for a 180'' diameter cloud (at a distance D = 160 pc)
with n$_0$ = 4 10$^5$ cm$^{-3}$, r$_0$ = 20" and a uniform abundance 
X(N$_2$H$^+$) = 2 10$^{-10}$. 
The velocity fields compared are linear functions of r with 
slopes of $\pm$ 1.5 km s$^{-1}$ pc$^{-1}$
(the velocity at the outer radius is $\pm$ 0.1 km s$^{-1}$) . 

In non-LTE conditions, the line excitation temperatures follow the 
density profile and, in the present case, decrease outward. Thus, in the case of 
a linearly decreasing velocity field, photons emitted at the center of the cloud 
appear blue-shifted for the molecules at greater radii which then absorb the red 
wing of the central emission (the opposite occurs for a linearly increasing 
velocity field). This effect is important if the lines are optically
thick. Thus, in typical interstellar clouds conditions, self absorption 
affects the $J$=2-1 transition (see figure \ref{f9} for the opacities) while the emerging 
profiles of the $J$=1-0 and 3-2 lines are similar with and without velocity fields.   

On Figure \ref{f11}, we see that the effect of the infalling material 
is to enhance the blue wing of the $J$=2-1 central component which is self absorbed 
in the static model, the red wing 
of the central component being reduced by self-absorption. In the 
case of a velocity field linearly increasing with r, the opposite occurs and emission in 
the red wing is enhanced. Note that because the $J$=2-1 central component
is composed of blended hyperfine lines with different line strengths 
(see figure \ref{f11}), changes in the line profile are not similar with respect 
to a change in the sign of the slope. 

Also, we stress that this effect is similar to what
happens to rotational lines with large opacity, except that, in the 
present case, the $J$=2-1 central component corresponds to hyperfine blended 
lines of small intrinsic line width. Thus, characteristic features appear even if 
the velocity gradient is weak. 

\subsection{Rate coefficients}

\begin{figure}
\includegraphics[angle=270,scale=0.6]{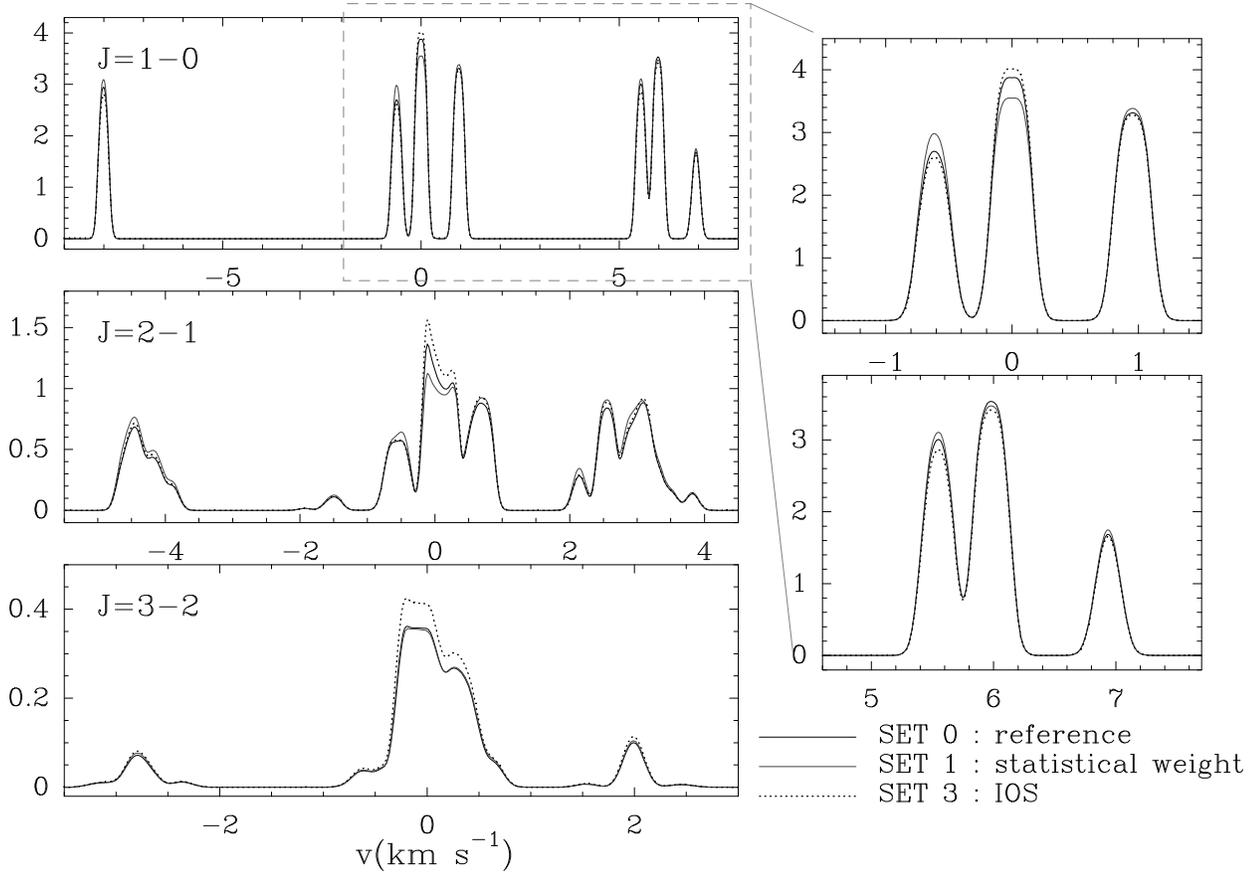}
\caption{
$J$=1--0, 2--1 and 3--2 emerging profiles (left column) for a cloud 
at T$_K$ = 10K, n(H$_2$) = 2 10$^5$ cm$^{-3}$ and X(N$_2$H$^+$) = 5 10$^{-10}$. 
Two approximations for the hyperfine collisional rate
coefficients (see text) are compared to the set of 
reference (black solid line). The first consist of rates proportional 
to statistical weights (gray solid lines) and the second is based on the 
IOS approximation (dashed lines). The right column corresponds to a blow--up
of the $J$=1--0 hyperfine lines.
\label{f12}}
\end{figure}

In star forming regions, the relatively large abundance of 
H$_2$ makes this species the main colliding partner for other 
molecules. Estimated values of the collisional rate coefficients 
of molecular species in collision with para-H$_2(j=0)$ 
can be obtained from the ones calculated with He as the collision 
partner. The underlying approximation is to consider identical cross 
sections values for the two colliding systems, and then apply a scaling 
factor of 1.37 to the rate coefficients in order to correct for the 
associated different reduced masses. However, by analogy with the 
results obtained on the collision systems involving HCO$^+$, one 
could expect the cross sections values with H$_2$ as a collision partner to 
be 2-3 times larger than those with He depending on the selected 
rotational transition. As discussed in section \ref{rate_LVG}, the main 
effect introduced by higher
collisional rate coefficients is to lower the critical
density of the rotational lines, which has direct consequences on the
determination of both density and molecular abundance. The hyperfine
rate coefficients are also changed through variations of the ratio
of the different P$^K_{jj'}$ \citep[see][]{dan04}. Indeed, these 
variations
modify the relative importance of hyperfine rate coefficients
among a given rotational line.
In order to assess the influence of such variations, we carried
out calculations where two distinct approximations are employed to 
determinate the hyperfine rate coefficients.
The first consisted on rate coefficients
proportional to the statistical weights of the final levels
of the transitions and the second was based on the IOS
approximation \citep[see][]{dan05}. Figure \ref{f12} shows the
emerging profiles resulting from the three sets of calculations for a 
cloud at T$_K$ = 10 K,
n(H$_2$) = 2 10$^5$ cm$^{-3}$ and X(N$_2$H$^+$) = 5 10$^{-10}$.
It is readily seen that the relative intensities
of $J$=1--0 hyperfine lines are similar and that the largest difference
($\sim$ 15-20\%) is encountered
for the thickest $JF_1F$ = 123--012 line. The three sets give the same
results for the thinnest $JF_1F$ = 110--011 line. Moreover, the 
intensity in the $J$=3--2
line associated with the IOS set of collisional rates is enhanced 
compared to the two other sets. This is related to the high
propensity of the $\Delta F = \Delta F_1 = \Delta J$ transitions which 
lead
to a more efficient pumping of the high--$J$ levels. Note that it 
entails higher
T$_{ex}$ for the $J$=2--1 hyperfine lines which reduce the 
self-absorption
feature in the main $J$=2--1 hyperfine component.
Finally, one would expect the collisional rate coefficients calculated with 
H$_2$
to affect similarly the intensities in the $J$=1--0 line, through
variations of the ratio of the opacity tensors P$^K_{jj'}$. Thus, as 
discussed
above, the most important effect introduced by rate coefficients 
calculated
with H$_2$ would be to scale all the rate coefficients and thus modify 
the density
and abundance estimates.

\section{Conclusions}\label{conclusion}

We have used numerical methods in order to investigate the excitation 
properties of molecules with a hyperfine energy structure, focusing specially 
on N$_2$H$^+$. Our conclusions are summarized as follows :
\begin{enumerate}
\item The hyperfine structure must be taken into account in the
radiative transfer calculations in order to derive
the total amount of molecules present in a given rotational level. 
When the energy structure of the molecule is restricted to its rotational 
energy structure, high-$J$ levels are more 
efficiently populated and the opacities of the low-$J$ transitions are underestimated. 
The error increases with the column density of N$_2$H$^+$. Errors as
large as a factor 2 could be induced if the hyperfine
structure of N$_2$H$^+$ is neglected.
\item For the typical temperature of dark clouds, i.e. T$_K$=10 K, and 
n(H$_2$) $\textless$ 10$^6$ cm$^{-3}$ 
the ratio of hyperfine brightness temperatures derived in the LVG 
approximation are always smaller than the ones predicted using the LTE  
approximation. Thus, this latter method is inadequate to assess
densities or column densities from observed N$_2$H$^+$ lines for the typical 
conditions prevailing in such clouds.
\item The assumption of equal excitation temperatures for all hyperfine components  
belonging to the same rotational transition is valid for low opacities and 
fails down for high opacities due to radiative processes.
For low opacities, we do not find any difference in the excitation
temperatures of the different hyperfine transitions induced
by collisional rate coefficients. Our calculations indicate that for
temperatures in the range 5-50 K and for all volume densities, the
excitation temperature of these lines will be identical if the total
opacity of the $J$=1-0 line is small. 
We stress that such collisional excitation effects, often found in the
literature, may not be invoked to
explain the intensity anomalies reported for N$_2$H$^+$.
\item Non-local radiative transfer results show that the $J$=1-0 intensity 
anomalies arise while the opacity of this line increases. Moreover, photon 
scattering in the low density envelopes affect the $J$=1-0 hyperfine lines 
differently and tend to reduce the emerging intensity of the thicker lines 
whereas the intensity of the thinnest lines is not modified by the envelope. 
Thus, in order to reproduce the intensities of the $J$=1-0 hyperfine lines, 
the only valid method is a non-local computation of the radiative transfer.  
\item For the typical conditions of cold dark clouds, the $J$=2-1 line is a 
good probe of low velocity fields. Nevertheless, this transition is difficult 
to observe from ground based observatories due to the high opacity of the 
atmosphere at 186 GHz, although some work can be carried out in dry 
weather conditions.
\end{enumerate}

\acknowledgments
The authors thank the European Union for support under the FP6 program
"The Molecular Universe" and the Spanish/French Picasso project
HF2003-0293. J. Cernicharo would like to thank Spanish MEC
for funding support through AYA2003-2785, AYA2000-1784,
ESP2001-4516, ESP2002-1627, ESP2002-12407, AYA2003-10113, and 
ESP2004-00665. Fabien Daniel wishes to thank Spanish MAE-AECI 2004/2005
for a fellowship grant from Program IIa. This study is partly supported 
by the European Community's human potential Program under contract 
MCRTN 512302, the "Molecular Universe".
We are grateful to A. Spieldfieldel, J.R. Pardo and F. Dayou for their help
and useful comments concerning the present manuscript.

\clearpage

\begin{table}[t]
\begin{tabular}{|c|c|c|c|c|c|}
\hline
($J$$F_1$F)$_u$ & ($J$$F_1$$F$)$_l$ & Frequency (MHz) & A$_{ul}$ (10$^{-5}$ s$^{-1}$) & s$_{ul}$ \\
\hline
 110 & 011 &   93171.617 &  3.628 &   0.333 \\
 112 & 011 &   93171.913 &  0.907 &   0.417 \\
 112 & 012 &   93171.913 &  2.721 &   1.250 \\
 111 & 010 &   93172.048 &  1.209 &   0.333 \\
 111 & 011 &   93172.048 &  0.907 &   0.250 \\
 111 & 012 &   93172.048 &  1.512 &   0.417 \\
 122 & 011 &   93173.475 &  2.721 &   1.250 \\
 122 & 012 &   93173.475 &  0.907 &   0.417 \\
 123 & 012 &   93173.772 &  3.628 &   2.330 \\
 121 & 010 &   93173.963 &  2.015 &   0.556 \\
 121 & 011 &   93173.963 &  1.512 &   0.417 \\
 121 & 012 &   93173.963 &  0.101 &   0.028 \\
 101 & 010 &   93176.261 &  0.403 &   0.111 \\
 101 & 011 &   93176.261 &  1.209 &   0.333 \\
 101 & 012 &   93176.261 &  2.016 &   0.556 \\
\hline
\end{tabular}
\caption{Frequencies, Einstein coefficients (A$_{ul}$) and line strengths (s$_{ul}$)
for the hyperfine components of the $J$=1-0 transition
of N$_2$H$^+$, as given by eq. \ref{coeff_hyp} (see text). Index $u$ ($l$) denote 
the initial (final) levels quantum numbers.\label{table:spectroscopy}}
\end{table}
\clearpage

\clearpage

\clearpage

\clearpage

\clearpage

\clearpage

\clearpage

\clearpage

\clearpage

\clearpage

\clearpage

\clearpage


\begin{thebibliography}{}
\bibitem[Belloche \& Andr\'e(2004)]{bel04}Belloche A., and Andr\'e P., 2004, A.\&A., 419, 35 
\bibitem[Bergin et al.(2002)]{ber02}Bergin E.A., Alves J., Huard T., Lada C.J., 2002, ApJ, 570, L101
\bibitem[Botschwina(1984)]{bot84} Botschwina P., 1984, Chem. Phys. Lett., 107, 535
\bibitem[Caselli et al.(1995)]{cas95}Caselli P., Myers P., Thaddeus P., 1995, ApJ, 455, L77
\bibitem[Caselli et al.(2002)]{cas02}Caselli P., Benson P.J., Myers P., Tafalla M., 2002, ApJ, 572, 238
\bibitem[Cernicharo \& Gu\'elin(1987)]{cer87}Cernicharo J., Gu\'elin M., 1987, A.\&A., 176, 299
\bibitem[Daniel et al.(2004)]{dan04} Daniel F., Dubernet M.-L. and Meuwly M., 2004, J. Chem. Phys., 121, 4540
\bibitem[Daniel et al.(2005)]{dan05} Daniel F., Dubernet M.-L., Meuwly M., Cernicharo J., 2005, MNRAS, 363, 1083
\bibitem[G\'onzalez-Alfonso \& Cernicharo(1993)]{gon93}G\'onzalez-Alfonso E., Cernicharo J., 1993, A.\&A., 279, 506
\bibitem[Gordy \& Cook(1984)]{gordy} Gordy W. and Cook R.L., \textit{Microwave molecular spectra}, Techniques of chemistry vol 18., Wiley Interscience Publication, 1984
\bibitem[Green(1975)]{gre75}Green S., 1975, ApJ, 201, 366
\bibitem[Havenith et al.(1990)]{hav90}Havenith M., Zwart E., Leo Meerts W., ter Meulen J.J., 1990, J. Chem. Phys., 93, 8446
\bibitem[Hotzel et al.(2004)]{hot04} Hotzel S., Harju J., Walmsley M.C., 2004, A.\&A., 415, 1065
\bibitem[Lee et al.(1999)]{lee99}Lee C.W., Myers P.C., Tafalla M., 1999, ApJ, 526, 788
\bibitem[Tafalla et al.(2004)]{taf04}Tafalla M., Myers P.C., Caselli P., Walmsley C.M., 2004, A.\&A., 416, 191
\bibitem[Thaddeus \& Turner(1975)]{tha75}Thaddeus P. and Turner B.E., 1975, ApJ, 201, L25
\bibitem[Turner \& Thaddeus(1977)]{tur77}Turner B.E. and Thaddeus P., 1977, ApJ, 211, 755
\bibitem[Ward-Thompson et al.(1994)]{war94} Ward-Thompson D., Scott P.F., Hills R.E., Andr\'e P., 1994, MNRAS, 268, 276

\end{thebibliography}
\end{document}